\documentclass[twocolumn,aps,prl,groupedaddress,showpacs]
              {revtex4}
              \usepackage{amssymb}%
\usepackage{graphicx}
\usepackage{amsmath}
\usepackage{bm}
\usepackage{relsize}
\RequirePackage{xspace}
\begin{document}

%\preprint{\begin{tabular}{l}
%          FERMILAB-PUB-02/332-T\\
%          ANL-HEP-PR-02-110\end{tabular}}

\title{\mbox{}\\[10pt]
Factorization and NLO QCD correction in $\bm{e^+ e^- \to J/\psi
(\psi(2S))+ \chi_{c0}}$ at B Factories}
% \altaffiliation{}

\author{Yu-Jie Zhang$~^{(a)}$, Yan-Qing Ma$~^{(a)}$, and Kuang-Ta Chao$~^{(a,b)}$}
\affiliation{ {\footnotesize (a)~Department of Physics and State Key
Laboratory of Nuclear Physics and Technology, Peking University,
 Beijing 100871, China}\\
{\footnotesize (b)~Center for High Energy Physics, Peking
University, Beijing 100871, China}}

%\email{ktchao@th.phy.pku.edu.cn}
%\thanks{}


%\date{\today}

%%%%%%%%%%%%%%%%%%%%%%%%%%%%%%%%%%%%%%%%%%%%%%%%%%%%%%%%%%%%%%%%%%%%%%%%%%%%%%
\begin{abstract}
In nonrelativistic QCD (NRQCD), we study %the double charmonium
%production
$e^+ e^- \to J/\psi(\psi(2S)) +\chi_{c0}$ at $B$ factories, where
the P-wave state $\chi_{c0}$ is associated with an S-wave state
$J/\psi$ or $\psi(2S)$. In contrast to the failure of factorization
in most cases involving P-wave states, e.g. in $B$ decays, we find
that factorization holds in this process at next to leading order
(NLO) in $\alpha_s$ and leading order (LO) in $v$, where the
associated S-wave state plays a crucial rule in canceling the
infrared (IR) divergences. We also give some general analyses for
factorization in various double charmonium production. The NLO
corrections in $e^+ e^- \to J/\psi(\psi(2S)) +\chi_{c0}$ at
$\sqrt{s}=10.6$~GeV are found to substantially enhance the cross
sections by a factor of about 2.8; hence crucially reduce the large
discrepancy between theory and experiment. With $m_c=1.5{\rm GeV}$
and $\mu=2m_c$, the NLO cross sections are estimated to be
$17.9(11.3)$~fb for $e^+ e^- \to J/\psi(\psi(2S))+ \chi_{c0}$, which
reach the lower bounds of experiment.
\end{abstract}
% 13.66.Bc   Hadron production in e-e+ interactions
% 12.38.Bx   Perturbative calculations
% 14.40.Gx   Mesons with S=C=B=0, mass > 2.5 GeV (including quarkonia)
%\keywords{}

\maketitle

The production of double charmonium in $e^+e^-$ annihilation at B
factories
\cite{Uglov:2004xa,Abe:2002rb,Abe:2004ww,cs,Braaten:2002fi,Liu:2002wq,
Hagiwara:2003cw,Bodwin:2002fk, BaBar:2005,Braguta:2006nf,
MaandSi:2004,bondar,bodwin06,he,Brambilla:2004wf} is one of the
challenging problems in heavy quarkonium physics and % nonrelativistic
%quantum chromodynamics (NRQCD)
NRQCD\cite{BBL}. For $e^+ e^- \to J/\psi \eta_c$ the QCD radiative
correction has turned out to be essential to greatly enhance the
theoretical prediction in NRQCD\cite{Zhang:2005cha,Gong:2007db}.
However, the cross sections of other processes, i.e., $e^+ e^- \to
J/\psi (\psi(2S))\chi_{c0}$ measured by Belle\cite{Uglov:2004xa}
%--------------
\begin{eqnarray}
\sigma[J/\psi +  \chi_{c0}] \times B^{ \chi_{c0}}[> 2] &=& \left( 16
\pm 5 \pm 4 \right) \; {\rm fb}, \label{Belle}\nonumber \\
\sigma[\psi(2S) +  \chi_{c0}] \times B^{ \chi_{c0}}[> 2] &=& \left(
17 \pm 8 \pm 7 \right) \; {\rm fb},
\end{eqnarray}
are also larger than LO NRQCD predictions by about an order of
magnitude or at least a factor of 5. Here $B^{ \chi_{c0} }[> 2 ]$ is
the branching fraction for the $\chi_{c0} $ decay into more than 2
charged tracks. Theoretically,  two studies in NRQCD by Braaten and
Lee\cite{Braaten:2002fi} and by Liu, He, and Chao\cite{Liu:2002wq}
showed that, at LO in the QCD coupling constant $\alpha_s$ and the
charm quark relative velocity $v$, the cross-section of $e^+ e^- \to
J/\psi (\psi(2S)) \chi_{c0}$ at $\sqrt{s}=10.6$GeV is only about
$2.4\sim 6.7(1.0\sim4.4)$fb (depending on the used parameters, e.g.,
the long-distance matrix elements, $m_c$ and $\alpha_s$).
%The cross-section is also
%calculated in the framework of light cone formalism
%\cite{Braguta:2006nf}.
So, it is crucial to verify that the QCD radiative correction can
also greatly enhance $\sigma[J/\psi (\psi(2S)) \chi_{c0}]$, before
we can claim that the large discrepancy between theory and
experiment for double charmonium production is really resolved.

However, we do need proof for the validation of factorization in
exclusive production processes involving P-wave states at NLO in
NRQCD. In fact,  nonfactorizable infrared (IR) divergences are found
in e.g. the P-wave charmonium production in $B$ meson exclusive
decays such as $B\to \chi_{c0}K$\cite{Song:2003yc}, in contrast to
the factorizable S-wave charmonium production in $B\to
J/\psi(\eta_c)K$\cite{cheng2001}, though the IR divergence in the
P-wave case is $m_c/m_b$ power suppressed\cite{Song:2003yc}. This
nonfactorizable feature for the P-wave states is a quite general
result  and is essentially due to the non-vanishing relative
momentum between the heavy quark and antiquark in P-wave states. So,
differing from the double S-wave charmonium production, where
factorization can be expected to generally hold at NLO in QCD, it is
crucial to prove the validation of factorization in the special
process $e^+ e^- \to J/\psi (\psi(2S))\chi_{c0}$, where the P-wave
charmonium is involved. Recently, the color transfer has been
noticed\cite{Nayak:2007mb} in associated heavy-quarkonium
production, e.g., $e^+ e^- \to J/\psi c \bar {c}$, where IR
divergence appears due to soft interactions between the associated
$c$ (or $\bar c$) quark and the $c\bar c$ pair of charmonium, and
hence breaks down factorization at NNLO. Therefore, it is
significant to clarify related problems in QCD for $e^+ e^- \to
J/\psi (\psi(2S)) \chi_{c0}$. Moreover, NLO QCD corrections are also
important in understanding heavy quarkonium production at hadron
colliders\cite{maltoni07}.

In  this paper we will prove the validation of factorization for
$e^+ e^- \to J/\psi \chi_{c0}$ at NLO in QCD, and calculate the
radiative corrections, while we have already found the NLO QCD
corrections to $e^+ e^- \to J/\psi \eta_c$\cite{Zhang:2005cha} and
$e^+  e^- \to J/\psi+c \bar c$\cite{Zhang:2006ay} to be large, and
increase the cross sections by a factor of about 2. All these are at
LO in $v$.

\begin{figure}
\includegraphics[width=8.5cm]{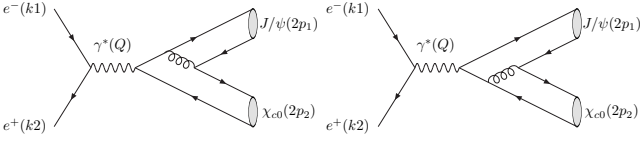}
\caption{\label{fig1}Two of four Born diagrams for $e^- e^+ \to
J/\psi \chi_{c0}$. }
\end{figure}

At LO in $\alpha_s$, $J/\psi + \chi_{c0}$ can be produced at order
$\alpha^2\alpha_s^2$, for which we refer to e.g.
Ref~\cite{Liu:2002wq}. The Feynman diagrams are shown in
Fig.~\ref{fig1}%, and the other two can be obtained by reversing the
%arrows on the quark lines
. Momenta for the involved particles are
assigned as $e^-(k_1) e^+ (k_2)\to J/\psi (2p_1)+ \chi_{c0} (2p_2)$.
In the calculation, we use {\tt FeynArts}~\cite{feynarts} to
generate Feynman diagrams and amplitudes, {\tt
FeynCalc}~\cite{Mertig:an} for the tensor reduction, and {\tt
LoopTools}~\cite{looptools} for the numerical evaluation of the
infrared (IR)-safe one-loop integrals.
%%%%%%%%%%%%%%%%%%%%%%%%%%%%%%%%%%%%%%%%%%%%%%%%%%%%%%%%%%%%%%%%%%%%%%%%%%%%%%
% Surround figure environment with turnpage environment for landscape
% figure
%\begin{widetext}
%\begin{center}
%\end{center}
%\end{widetext}

%%%%%%%%%%%%%%%%%%%%%%%%%%%%%%%%%%%%%%%%%%%%%%%%%%%%%%%%%%%%%%%%%%%%%%%%%%%%%%
% Surround figure environment with turnpage environment for landscape
% figure
%\begin{widetext}
%\begin{center}
\begin{figure*}
\includegraphics[width=14.5cm]{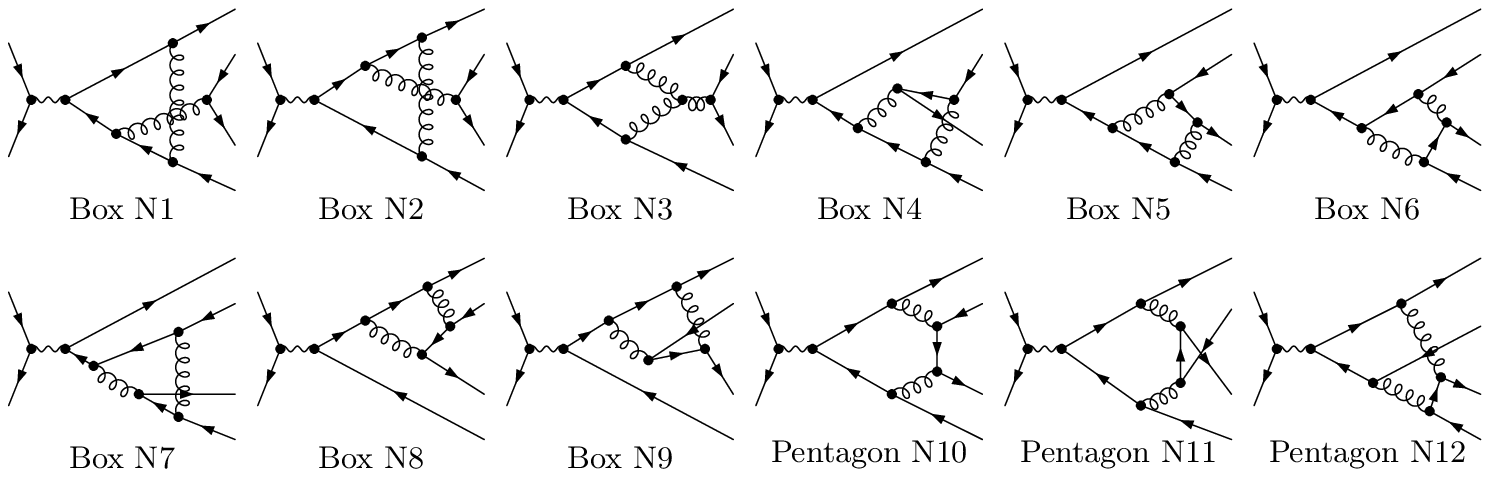}
\caption{\label{fig2} Twelve of the twenty-four box and pentagon
diagrams for $e^-(k_1) e^+ (k_2)\to J/\psi(2p_1) \chi_{c0}(2p_2)$.
Two upper charm legs are for $J/\psi$ while two lower ones for
$\chi_{c0}$.}
\end{figure*}
%\end{center}
%\end{widetext}
At NLO in $\alpha_s$, there are ultraviolet(UV), IR, and Coulomb
singularities. We choose renormalization schemes the same as in
\cite{Zhang:2005cha}, and use $D=4-2\epsilon$ dimension and relative
velocity $v$ to regularize IR and Coulomb singularities. For the box
diagrams shown in Fig.~\ref{fig2},Box N5, N8, N10 have IR and
Coulomb singularities, Box N3  is IR finite, while the other nine
diagrams have IR singularities. There are some scalar functions that
should be calculated by hand, which are usually defined as
\begin{equation}
T^i_0[p_1\ldots p_{i-1},m_0 \ldots m_{i-1}]%\nonumber \\&=&
=\int \frac{\mu^{2\epsilon }d^D
q/(2\pi)^D}{\prod_{l=0}^{i-1}[(q+p_l)^2-m_l^2]},\nonumber
\end{equation}
where $p_0=0$, and $T^i=C,D,E$ for $i=3,4,5$ respectively. Some of
them were given in \cite{Zhang:2005cha}. The others are
\begin{eqnarray}
\label{Ezero} &&\hspace{-0.7cm} \left. \frac{\partial
E_0[p_2\hspace{-0.1cm}+\hspace{-0.1cm} q,p_1\hspace{-0.1cm}+
\hspace{-0.1cm}p_2\hspace{-0.1cm}+\hspace{-0.1cm}q,
\hspace{-0.1cm}2p_1\hspace{-0.1cm}+\hspace{-0.1cm}p_2\hspace{-0.1cm}+\hspace{-0.1cm}q,
q\hspace{-0.1cm}-\hspace{-0.1cm}p_2,\hspace{-0.05cm}0\hspace{-0.05cm}
,\hspace{-0.05cm}m\hspace{-0.05cm},\hspace{-0.05cm}0\hspace{-0.05cm},m,m]}{\partial
q^\alpha}
 \right|_{q=0}\nonumber \\
&=&\hspace{-0.2cm}\left\{\frac{i\  }{24 m^4 s^2 \pi ^2} \left[4 (5-2
r) \sqrt{4-r} \log
\left(\frac{2-\sqrt{4-r}}{\sqrt{4-r}+2}\right)+\right.
\right.\nonumber \\
&&\left.\hspace{-0.2cm} \left.(10 \hspace{-0.1cm}+\hspace{-0.1cm}
18\log 2) r \hspace{-0.1cm}+\hspace{-0.1cm} (40
\hspace{-0.1cm}-\hspace{-0.1cm}18 r) \log\hspace{-0.07cm}
\left(\hspace{-0.07cm}
\frac{16}{r}\hspace{-0.07cm}\right)\hspace{-0.07cm}
\right]\hspace{-0.1cm}-\hspace{-0.1cm} \frac{192 C_0}
{s^3}\right\}p_1^\alpha,
\nonumber
\\
&&\hspace{-0.7cm}\left. \frac{\partial
D_0[p_2\hspace{-0.1cm}+\hspace{-0.1cm} q,p_1\hspace{-0.1cm}+
\hspace{-0.1cm}p_2\hspace{-0.1cm}+\hspace{-0.1cm}q,
q\hspace{-0.1cm}-\hspace{-0.1cm}p_2,0,m,0,m]}{\partial q^\alpha}
 \right|_{q=0}\nonumber \\
&=&\left[\frac{i (2+\log 4)}{m^2 \pi ^2 s^2}-\frac{32
C_0}{s^2}\right]p_1^\alpha,\nonumber \\
&&\hspace{-0.8cm} \left. \frac{\partial D_0[p_1,p_1\hspace{-0.1cm}+
\hspace{-0.1cm}p_2\hspace{-0.1cm}+\hspace{-0.1cm}q,
-\hspace{-0.1cm}p_1,0,m,0,m]}{\partial q^\alpha} \right|_{q=0}
\hspace{-0.1cm}\hspace{-0.3cm}=\hspace{-0.1cm}\frac{ i p_1^\alpha
\log 4}{m^2 \pi ^2 s^2}\hspace{-0.1cm}-\hspace{-0.1cm}\frac{32
p_1^\alpha C_0}{s^2}.\nonumber
\end{eqnarray}
Here $q$ is the relative momentum of charm quark in $\chi_{c0}$,
$r=16m^2/s$, and $C_0$ is the Coulomb and IR divergent three point
function $C_0[p_{c},-p_{\bar{c}},0,m,m]$%and $p_{c}, p_{\bar{c}}$
%are the momenta of charm and anti-charm quark in charmonium. It
\cite{Zhang:2005cha},
%With $v=|\overrightarrow{p_{c}}-
%\overrightarrow{p_{\bar c}}|/m \to 0$, we have
\begin{eqnarray}
C_0= \frac{-i}{2m^2(4\pi)^2}\left(\frac{4 \pi
\mu^2}{m^2}\right)^{\epsilon}\Gamma(1+\epsilon)\left[\,
\frac{1}{\epsilon} + \frac{\pi^2}{v} -2 \right].
\end{eqnarray}
The IR terms of $\mathrm{Box N5\hspace{-0.05cm}+\hspace{-0.05cm}N8
\hspace{-0.05cm}+\hspace{-0.05cm}Pentagon N10}$ are canceled by
counter terms, and the Coulomb singularity is mapped into the wave
functions. Other IR terms can be separated into three point
functions $C_0[p_{2}+q,-p_1]$ and
$C_0[p_{2}-q,-p_1]$\cite{Dittmaier:2003bc}. And we have
\begin{eqnarray}
\hspace{-0.3cm}  C_0[p_{2}+q,-p_1]\big|_{q=0}&&\hspace{-0.3cm}=
C_0[p_{2}-q,-p_1]\big|_{q=0},\\
\hspace{-0.3cm}\left. \frac{\partial C_0[p_{2}+q,-p_1] }{\partial
q^\alpha} \right|_{q=0}&&\hspace{-0.3cm}=-\left. \frac{\partial
C_0[p_{2}-q,-p_1] }{\partial q^\alpha} \right|_{q=0}.
\end{eqnarray}
Here  $C_0[l',-l]$ means $C_0[l',-l,0,m,m]$. %With these two
%equations,
Then we get that $\mathrm{Box N1 \hspace{-0.05cm} +\hspace{-0.05cm}
N4}$ and $\mathrm{Box N6\hspace{-0.05cm} +\hspace{-0.05cm} N7
+\hspace{-0.05cm} Pentagon N12} $ are IR finite. The IR terms of
$\mathrm{ Box
N9\hspace{-0.05cm}+\hspace{-0.05cm}N2\hspace{-0.05cm}+\hspace{-0.05cm}Pentagon
N11}$ are canceled by vertex diagrams. The UV terms are canceled by
counter terms. Then the final NLO result for the cross section is
UV-, IR-, and Coulomb-finite. Details of the calculation can be
found in a forthcoming paper. Since the result is IR finite, $e^+ +
e^-\rightarrow J/\psi+\chi_{c0}$ is factorizable at NLO in NRQCD
factorization.

%%%%%%%%%%%%%%%%%%%%%%%%%%%%%%%%%%%%%%%%%%%%%%%%%%%%%%%%%%%%%%%%%%%%%%%%%%%%%%
% Surround figure environment with turnpage environment for landscape
% figure
%\begin{widetext}
%\begin{center}
\begin{figure}
\includegraphics[width=6.6cm]{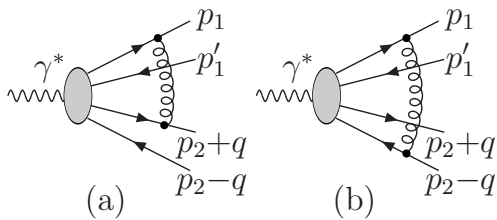}
\caption{\label{fig:factor}Half of the diagrams for one-loop virtual
IR corrections with two charm quark pairs $c(p_1) \bar c (p_1')$ and
$c(p_2+q) \bar c (p_2-q)$. The other two diagrams can be obtained by
replacing $c(p_1)$ with $\bar c (p_1')$.  }
\end{figure}
%\end{center}
%\end{widetext}

The key part for the cancelation of IR divergence in our calculation
is shown in Fig.~\ref{fig:factor}. The IR term of the NLO vertex
correction for two charm quarks of momenta $p_1$ and $p_2+q$ shown
in Fig.~\ref{fig:factor}(a) is proportional to $p_1\cdot (p_2+q)
C_0[-p_1,p_2+q_1]$ (see also \cite{Nayak:2007mb,Dittmaier:2003bc}
though the three-point function $C_0$ was not explicitly written
there); while it is proportional to $-p_1\cdot (p_2-q)
C_0[-p_1,p_2-q]$ for a charm quark and an anti-charm quark of
momentum $p_1$ and $p_2-q$ as shown in Fig.~\ref{fig:factor}(b).
Then for the charm quark $c(p_1)$ associated with a colorless charm
quark pair $c(p_2+q)$$\bar c (p_2-q)$,
% that was shown in Fig.~\ref{fig:factor}
the IR term becomes
\begin{eqnarray}
 &&\hspace{-0.1cm} {\mathcal{M}^{IR}_{NLO}[c(p_1)+c(p_2+q)\bar
c(p_2-q)]}
 \nonumber\\ \hspace{-0.5cm} \propto&&\hspace{-0.4cm}
  (p_2\hspace{-0.075cm}+\hspace{-0.075cm}q)\cdot p_1
C_0[p_2\hspace{-0.075cm}+\hspace{-0.075cm}q,-\hspace{-0.075cm}p_1]
-(p_2\hspace{-0.075cm}-\hspace{-0.075cm}q)\cdot p_1
C_0[p_2\hspace{-0.075cm}-\hspace{-0.075cm}q,-\hspace{-0.075cm}p_1]
\nonumber
\\
\label{eq:cccIR}\hspace{-0.5cm}=
%\propto
&& \hspace{-0.4cm}
%0\hspace{-0.075cm} +\hspace{-0.075cm} 2
 2q^\alpha
\hspace{-0.08cm}\Big[\hspace{-0.08cm} p_{1\alpha}
C_0[p_2,\hspace{-0.05cm}-p_1]\hspace{-0.075cm}  + \hspace{-0.075cm}
p_2 \hspace{-0.075cm}\cdot\hspace{-0.075cm} p_1\frac{\partial
C_0[\hspace{-0.08cm}p_2\hspace{-0.125cm}+\hspace{-0.075cm}q,\hspace{-0.05cm}-\hspace{-0.05cm}p_1]}{\partial
q^\alpha}\big|_{q=0} \hspace{-0.08cm}\Big] %\nonumber
\hspace{-0.075cm}+\hspace{-0.075cm}{\cal{O}}\hspace{-0.015cm}(\hspace{-0.06cm}q^2\hspace{-0.06cm}
)\hspace{-0.05cm},
\end{eqnarray}
where it is expanded in powers of the relative momentum $q$ at
$q=0$. (The IR terms between $c(p_2+q)$ and $\bar c (p_2-q)$ or
between $c(p_1)$ and $\bar c (p_1')$ are ignored since these $c\bar
c$ pairs should evolve to bound states at large distances.)

Following implications can be found from Eq.(\ref{eq:cccIR}):

(1) The IR term is finite when an associated charm quark connects
with both legs of the S-wave state $J/\psi$, where $q=0$ can be
taken at LO in $v$, and this corresponds to the zeroth order i.e.
the vanishing ${\cal{O}}(q^0 )$ term in Eq.(\ref{eq:cccIR}).

(2) The IR term becomes divergent when the associated charm quark
connects with both legs of the P-wave state $\chi_{c0}$, where the
relative momentum $q$ has to be retained, and this corresponds to
the first order i.e. the ${\cal{O}}(q^1 )$ term in
Eq.(\ref{eq:cccIR}),
%\begin{eqnarray}
% &&\hspace{-0.2cm} {\mathcal{M}^{IR}_{NLO}[c(p_1)+c(p_2+q)\bar
%c(p_2-q)]}
% \nonumber\\
%\label{eq:cccIRPP}\hspace{-0.5cm} \propto &&\hspace{-0.4cm}
%\left({\mathcal{M}_{Born}}\big|_{q=0}\right) \hspace{-0.04cm}
%\frac{8 q\cdot
%p_1}{s}\frac{\alpha_s}{\varepsilon_{IR}}\Big[\hspace{-0.04cm}
%1+{\cal{O}}\left(\frac{ m^2}{s}\right)\Big]
%\hspace{-0.075cm}+\hspace{-0.075cm}{\cal{O}}(q^2\hspace{-0.04cm}
%)\hspace{-0.04cm}.
%\end{eqnarray}
\begin{eqnarray}
 &&\hspace{-0.2cm} \mathcal{M}^{IR}_{NLO}
\label{eq:cccIRPP}\hspace{-0.05cm} \propto  \hspace{-0.04cm} \frac{8
q\cdot p_1}{s}\frac{\alpha_s}{\varepsilon_{IR}}\Big[\hspace{-0.04cm}
1+{\cal{O}}\left(\frac{ m^2}{s}\right)\Big]
\hspace{-0.075cm}+\hspace{-0.075cm}{\cal{O}}(q^2\hspace{-0.04cm}
)\hspace{-0.04cm}.
\end{eqnarray}
This nonvanishing IR divergence, which is actually independent of
the associated quark flavor, is the origin for the
non-factorizability in many processes involving P-wave states, e.g.,
in $B\to \chi_{c0}K$ decay\cite{Song:2003yc}, and also in $B$ decays
$B\to M_1M_2$ with $M_2$ being an emitted P-wave light meson (
$f_0$, $a_1$, $b_1$...)~\cite{Cheng} when the light meson mass e.g.
$m_{f_0}/m_B$ is not ignored in the IR divergent vertex corrections.
In another words, factorization holds up only to terms that are
$m_q/m_b$ power suppressed.

(3) When the associated fermion is an anti-quark $\bar c (p_1)$, we
can get a similar IR divergence by replacing $p_1$ with $-p_1$ in
Eq.(\ref{eq:cccIR}) and Eq.(\ref{eq:cccIRPP}). Then by adding
together the contributions of the associated charm pair $c (p_1)$
and $\bar c (p_1)$ connected with $c(p_2+q)\bar c (p_2-q)$, the IR
divergence is canceled in the case of P-wave e.g. $\chi_{c0}$.
%Furthermore,
%because the IR divergence in the amplitude is canceled in both real
%and imaginary parts at NLO in $\alpha_s$, the cross section of
%$e^+e^- \to J/\psi \chi_{cJ}$ should be factorizable at NNLO, in
%contrast to the case of $e^+e^- \to J/\psi c\bar c$ where
%factorization is expected to fail at NNLO for the cross
%section\cite{Nayak:2007mb}.
Note that since generally the associated quark pair $c (p_1)$$\bar c
(p_1')$ has $p_1\neq p_1'$ as shown in Fig.~\ref{fig:factor}, the IR
cancelation  is incomplete and the divergence still remains. This
means that factorization in $e^+e^- \to J/\psi \chi_{cJ}$ can hold
at LO in $v$ for the $J/\psi$ (i.e. $p_1=p_1'$) but not hold at NLO
in $v$ ($p_1\neq p_1'$). Based on Eq.(\ref{eq:cccIR}) we can draw a
general conclusion that the double charmonium (including all
$S,P,D,...$ wave states) production in $e^+e^- $ annihilation is
factorizable at NLO in $\alpha_s$ only on condition that one of the
double charmonium is an S-wave state in which the quark relative
momentum is ignored. Or, factorization holds up only to terms that
are $m_c^2/s$ power suppressed. A similar conclusion is also
obtained recently in a more general analysis for quarkonium
production\cite{Bodwin:2008nf}.
%So, the cancelation of IR
%divergencies for a P-wave charmonium depends on the existence of an
%associated S-wave $c\bar c$ state at LO in $v$.

\begin{figure}
\includegraphics[width=8.0cm]{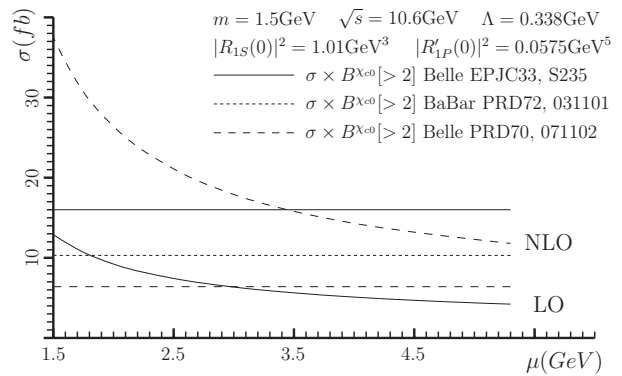}
\caption{\label{depmu}Cross sections of  $e^+ e^-\to J/\psi
+\chi_{c0}$  as functions of the renormalization scale $\mu$.  }
\end{figure}

We now turn to numerical calculations for the cross sections of $e^+
e^-\to J/\psi(\psi(2S))\chi_{c0}$. To be consistent with the NLO
result the values of wave functions squared at the origin should be
extracted from the leptonic width of $J/\psi(\psi(2S))$ and the
two-photon width of $\chi_{c0}$ at NLO in $\alpha_s$ (see
~\cite{BBL} and \cite{PDG}), we obtain $|R_{1S}(0)|^2 = 1.01 {\rm
GeV}^3$, $|R_{2S}(0)|^2 = 0.639 {\rm GeV}^3$, $|R'_{1P}(0)|^2=0.0575
{\rm GeV}^5$. Taking $m_{J/\psi}\!=\!m_{\psi(2S)}\!=\!
m_{\chi_{c0}}\!=\!2m$ at LO in $v$, $m\!=\!1.5$~GeV,
$\Lambda^{(4)}_{\overline{\rm MS}}\!=\!338{\rm MeV}$, we find
$\alpha_s(\mu)\!=\!0.259$ for $\mu\!=\!2m$, and get the cross
sections at NLO in $\alpha_s$
\begin{eqnarray}
\label{jsetac} \sigma(e^+ + e^-\rightarrow J/\psi
+\chi_{c0})&=&17.9 \rm{fb},\nonumber \\
\sigma(e^+ + e^-\rightarrow \psi(2S)+\chi_{c0})&=&11.3 \rm{fb},
\end{eqnarray}
which are a factor of $2.8$ larger than the LO cross sections
$6.35(4.02)$~fb for $J/\psi(\psi(2S))$. If we use the
BLM scale\cite{Brodsky:1982gc}, we get $\mu_{BLM}%=e^{-5/6}\sqrt s/2
=2.30{\rm GeV}$, $\alpha_s=0.291$, and the corresponding cross
sections $8.02(5.08)$~fb at LO and $22.8(14.4)$~fb at NLO.
Fig.~\ref{depmu} shows the cross sections at LO and NLO as functions
of the renormalization scale $\mu$, as compared with the Belle and
BaBar
data. Our LO and NLO results compared with experimental and other theoretical
cross sections are shown in Table I.
% In Fig.~\ref{depmcmc} we show the cross sections at LO and NLO
%as functions of the charm quark mass, as compared with the Belle and
%BaBar data.
We see the NLO QCD correction enhances the cross
sections by about a factor of 2.8, despite of existing theoretical
uncertainties.
%%%%%%%%%%%%%%%%%%%%%%%%%%%%%%%%%%%%%%%%%%%%%%%%%%%%%%%%%%%%%%%%%%%%%%%%%%%%%%
% Surround figure environment with turnpage environment for landscape
% figure
%\begin{widetext}
%\begin{center}
%\end{center}
%\end{widetext}

%%
%%%%%%%%%%%%%%%%%%%%%%%%%%%%%%%%%%%%%%%%%%%%%%%%%%%%%%%%%%%%%%%%%%%%%%%%%%%%%%%
%% Surround figure environment with turnpage environment for landscape
%% figure
%% \begin{turnpage}
%%\begin{widetext}
%%\begin{center}
%\begin{figure}
%\includegraphics[width=8.0cm]{psichiDEPMASS.eps}
%\caption{\label{depmcmc}Cross sections of  $e^+ e^-\to J/\psi
%+\chi_{c0}$ as functions of the charm quark mass.  }
%\end{figure}
%%\end{center}
%%\end{widetext}
%% \end{turnpage}
%%%%%%%%%%%%%%%%%%%%%%%%%%%%%%%%%%%%%%%%%%%%%%%%%%%%%%%%%%%%%%%%%%%%%%%%%%%%%%%
%

\begin{table}
\caption{\label{detalNo}Experimental and theoretical cross sections
of $e^+ e^-\to J/\psi (\psi(2S))\chi_{c0}$ at B factories in units
of fb. We use $|R_{1S}(0)|^2=1.01 {\rm GeV}^3$, $|R_{2S}(0)|^2=0.639
{\rm GeV}^3$, $|R'_{1P}(0)|^2=0.0575 {\rm GeV}^5$, $\Lambda=0.338
{\rm GeV}$, $\sqrt s=10.6{\rm GeV} $, $m_c=1.5{\rm GeV}$, and
$\mu=2m_c$. The experimental data are the cross sections
%of $J/\psi(\psi(2S)) +\chi_{c0}$
times the branching fraction for
$\chi_{c0} $ decay into more than 2 charged tracks. But the Belle
data of $\psi(2S) +\chi_{c0}$ in Ref.\cite{Abe:2004ww} correspond to
$\chi_{c0} $ decay into at least 1 charged tracks. }
\begin{ruledtabular}
\begin{tabular}{c|cc}
&  $J/\psi +\chi_{c0}$  &  $\psi(2S) +\chi_{c0}$ \\
\hline Belle $\sigma \times{ B^{ \chi_{c0}}[> 2]
}$\cite{Uglov:2004xa} & $16 \pm 5 \pm 4 $&  $17 \pm 8 \pm 7 $
\\
Belle $\sigma \times{ B^{ \chi_{c0}}[> 2(0)] }$\cite{Abe:2004ww} &
$6.4 \pm 1.7 \pm 1.0 $&  $12.5 \pm 3.8 \pm 3.1
$\\
BaBar $\sigma \times{ B^{ \chi_{c0}}[> 2] }$\cite{BaBar:2005}
& $10.3 \pm 2.5 ^{+1.4}_{-1.8} $&- \\
Braaten and Lee \cite{Braaten:2002fi}& 2.4& 1.0\\
Liu, He and Chao \cite{Liu:2002wq}& 6.7& 4.4 \\
Braguta et al. \cite{Braguta:2006nf} &14.4& 7.8\\
Our LO result & 6.35 & 4.02 \\
Our NLO result & 17.9 & 11.3
\end{tabular}
\end{ruledtabular}
\end{table}

We emphasize again the crucial rule of the associated S-wave state
$J/\psi$ payed in canceling the IR divergencies in the vertex
corrections in $e^+ e^-\to J/\psi\chi_{c0}$.   At LO in $v$ and NLO
in $\alpha_s$, the interaction of $\chi_{c0}$ with the charm quark
(or antiquark) in the $J/\psi$ is individually IR divergent, but the
sum of that of the charm quark and antiquark in the $J/\psi$ becomes
IR finite.  This result reflects the fact that the P-wave state
$\chi_{c0}$ behaves as a color dipole, which interacts with the
color charge carried by the charm quark (or antiquark) in the
$J/\psi$, but the interactions vanish when the charm quark and
antiquark in the $J/\psi$ are combined into a colorless S-wave
object at LO in $v$ (see  Eq.(\ref{eq:cccIR})). The validation of
factorization at NLO for $e^+ e^-\to J/\psi\chi_{c0}$ depends
crucially on the associated S-wave state $J/\psi$.

In conclusion, we find that at NLO in $\alpha_s$ and LO in $v$,
NRQCD factorization holds for the double charmonium production $e^+
e^- \to J/\psi(\psi(2S)) \chi_{c0}$.  We get UV and IR finite
corrections to the cross sections at $\sqrt{s}=10.6$~GeV, and the
NLO QCD corrections can substantially enhance the cross sections
with a K factor (the ratio of NLO to LO ) of about 2.8; and hence it
crucially reduces the large discrepancy between theory and
experiment. With $m=1.5{\rm GeV}$ and $\mu=2m$,
the NLO cross sections are estimated to be $17.9(11.3)$~fb, %for
%$J/\psi(\psi(2S))$
 which reach the lower bounds of experiment.

\begin{acknowledgments}
We thank G.T. Bodwin,  Y. Jia, J.P. Ma and J.W. Qiu for helpful
comments and discussions. This work was supported by the National
Natural Science Foundation of China (No 10675003, No 10721063), and
also by China Postdoctoral Science Foundation (No 20070420011).
\end{acknowledgments}


\end{document}